\documentclass[traditabstract]{aa} 
\usepackage{graphicx}
\usepackage{txfonts}
\usepackage{subfigure}
\usepackage{multirow}
\usepackage{natbib}
\usepackage{appendix}
\usepackage{array}
\usepackage{soul}
\usepackage[normalem]{ulem}

\usepackage{wasysym}

\newcommand{\gtap}{\mathrel{\hbox{\rlap{\lower.55ex \hbox {$\sim$}}
                   \kern-.3em \raise.4ex \hbox{$>$}}}}
\newcommand{\ltap}{\mathrel{\hbox{\rlap{\lower.55ex \hbox {$\sim$}}
                   \kern-.3em \raise.4ex \hbox{$<$}}}}

\begin{document}
  \title{A phenomenological model for the X-ray spectrum of Nova V2491 Cygni}

   \author{C. Pinto
          \inst{1,3}
          \and
          J.-U. Ness\inst{2}
          \and
          F. Verbunt\inst{1,3}
          \and
          J. S. Kaastra\inst{1,3}
          \and
          E. Costantini\inst{1}
          \and
          R. G. Detmers\inst{1}
          }

   \institute{SRON Netherlands Institute for Space Research,
              Sorbonnelaan 2, 3584 CA Utrecht, The Netherlands\\
              \email{c.pinto@sron.nl}
         \and
             ESAC European Space Astronomy Center, P.O. Box 78, 28691 Villanueva de la Canada, Madrid, Spain 
         \and
             Department of Astrophysics/IMAPP, Radboud University, PO BOX 9010, 6500 GL Nijmegen, The Netherlands\\
             }

   \date{Received August 9, 2011; Accepted June 10, 2012}

 
 \abstract{The X-ray flux of Nova V2491\,Cyg reached a maximum some forty days after optical maximum. The X-ray spectrum at that time, obtained with the RGS of XMM-Newton, shows deep, blue-shifted absorption by ions of a wide range of ionization. We show that {the deep absorption lines of the} X-ray spectrum at maximum, and nine days later, are well described by {the following phenomenological} model with emission from a central blackbody and from a collisionally ionized plasma (CIE). The blackbody spectrum (BB) is absorbed by three main highly-ionized expanding shells; the CIE and BB are absorbed by cold circumstellar and interstellar matter that includes dust. The outflow density does not decrease monotonically with distance. The abundances of the shells indicate that they were ejected from an O-Ne white dwarf. We show that the variations on time scales of hours in the X-ray spectrum are caused by a combination of variation in the central source and in the column density of the ionized shells. Our {phenomenological} model gives the best description {so far} of the supersoft X-ray spectrum {of nova V2491 Cyg}, but underpredicts, by a large factor, the optical and ultraviolet flux. The X-ray part of the spectrum must originate from a very different layer in the expanding envelope, presumably much closer to the white dwarf than the layers responsible for the optical/ultraviolet spectrum. This is confirmed by absence of any correlation between the X-ray and UV/optical observed fluxes.}


   \maketitle
%

\section{Introduction}

The sensitivity to soft X-rays of  the EXOSAT satellite enabled
the first discovery of X-rays from a classical nova, GQ\,Mus, in April
1984 \citep{Oegelman1984}. The ROSAT satellite detected it almost
eight years later, in Feb 1992, and found it to have a very soft blackbody-like spectrum with
$kT_{\rm eff}\sim28$\,eV
\citep{Oegelman1993}. ROSAT observations of other novae in our
galaxy showed that only a small fraction of novae are detected
in X-rays \citep{Oegelman1993, Orio2001}.

The spectrum of \object{GQ Mus} is very soft, comparable to the spectra of a
new class of supersoft X-ray sources (SSS) discovered with ROSAT in the
Large Magellanic Cloud, with temperatures $\ltap100$ eV in blackbody
fits to their spectra \citep{Truemper1991, Greiner1991}.
The favored explanation for the X-ray spectra of the supersoft
sources, as for GQ\,Mus, is hydrogen burning at the surface of a white
dwarf, of material accreted from a companion star. The class of
supersoft X-ray sources shows a large variety, some sources being
permanent and others transient, and with different types of 
companion stars \citep[see review by][]{Kahabka1997}.
Observation campaigns with ROSAT, XMM-Newton and Chandra in M31 show
that most of the supersoft sources are classical novae \citep{Pietsch2005, Henze2010}. 
{Recently, \citet{Henze2011} argued that that the high fraction of novae without detected SSS emission 
might be explained by the incomplete coverage. }

The model fits of the supersoft X-ray spectra have increased in sophistication
over the years. Black body fits indicated highly super-Eddington
luminosities for the ROSAT sources.  \citet{Heise1994}
showed that the fit of local thermodynamic equilibrium (LTE) model atmospheres to the ROSAT data brought
their bolometric luminosity below the Eddington limit, for
temperatures $\leq60$\,eV. For temperatures $\geq$\,100eV, \citet{Hartmann1997} showed that the differences between bolometric
luminosities for ROSAT supersoft sources derived from blackbody, LTE
and NLTE models are much smaller, the main difference being that NLTE
models may give appreciable flux above 0.4\,keV (i.e.\ in the `hard'
ROSAT band). A new level of sophistication is reached by \citet{Petz2005}
 who fit NLTE model spectra of expanding atmospheres, thus
taking into account the outflow inherent in the nova phenomenon.
Even the most sophisticated model fails to fit
the observed continuum of the nova \object{V4743 Sgr} at $\lambda<30$\,\AA,
and also does not match the depth of the absorption lines
\citep[Fig.\,3 of][]{Petz2005, Rauch2010}. 
{ With regard to the X-ray spectra of V2491 Cyg,  the best published 
 atmosphere model does not reproduce the observed line velocities
 and does not reproduce the observed line depths, as clearly visible
 in Fig.~11 of Ness et al.~(2011), for example in the \ion{N}{vii} 1s-2p line
 at 24.8~\AA. Atmosphere models so far also 
 fail to reproduce the wide range of ionization levels simultaneously
 present in the observed spectra: e.g. \ion{O}{v} to \ion{O}{viii} in V2491 Cyg
 (see Figs.~\ref{fig:rgs_obs1}-\ref{fig:rgs_obs4} below).}

What is the reason for this failure?  If there are shocks in the
expanding nova envelope, the assumption of a standard atmosphere that
density decreases monotonically outwards, breaks down. The rapid X-ray
variability observed in some novae indicates that the expanding
envelope and/or the luminosity from the deeper layers are not
stationary and/or not spherically symmetric, invalidating another
assumption in atmosphere models.  In this paper we therefore
investigate a pilot model of a rather different nature, in which
separate expanding shells absorb the spectrum of an underlying central
source.  In this first study, we  describe the central source as a
blackbody spectrum.  We apply our simple model to the XMM-Newton
RGS spectra obtained from the nova V2491\,Cyg to investigate whether
this model gives a better description of the continuum and line
absorption; whether the short-term variability is due to changes
in the source or in the absorption; and whether the model allows
determination of the chemical abundances in the expanding envelope.

Nova \object{V2491 Cyg} was discovered on April 10, 2008, at $V=7.7$ \citep{Nakano08}.
The optical flux declined from this peak in the
course of the following months, interrupted by a brief second maximum
at the end of April.  Ultraviolet and X-ray fluxes of the V2491\,Cyg
were obtained with the Swift satellite: the ultraviolet flux declined
in tandem with the optical flux (measured through May and June),
whereas the X-ray luminosity rose from a marginal detection on April
11 to a peak more than a thousand times brighter some 40 days later
\citep{Page10}. Lightcurves of optical, ultraviolet and X-ray are
shown in Fig.\,4 of \citet{Page10}.  In the Swift X-ray range
(0.3-8 keV) the spectrum peaks strongly near 0.4-0.5 keV during the
X-ray maximum and has a flatter (i.e.\,harder) energy distribution
before and after X-ray maximum \citep[see Fig.\,3 of][]{Page10}.

From a relation between maximum magnitude of novae and their rate of
decline, a distance of 10.5\,kpc has been estimated for V2491\,Cyg by
\citet{Helton08}. The optical flux was already declining during
the first measurements, and therefore the actual flux maximum may have
been higher than the flux of the first measurement: this would imply a
smaller distance.

The interstellar reddening $E(B-V)=0.43$ has been estimated from the
ratio of the O I lines at 0.84 and 1.13 micron \citep{Rudy08}, and
may be converted to a hydrogen column $N_H=2.4\times
10^{25}\mathrm{m}^2$ using the relation determined by
\citet{Predehl1995}.

The X-ray spectra described in this paper were obtained with 
XMM-Newton just before the X-ray maximum, and some nine days later,
when the X-ray flux had declined by an order of magnitude. 

Our paper is structured as follows. 
In Section\,2 we describe the XMM-Newton data and provide a brief
description of the four spectra that we will analyze.
In Section\,3 we describe a basic model and apply it to the spectra.
In Section\,4 we discuss several variant models and their applications.
In Section\,5 we discuss our model fits and in Section\,6 we summarize
our conclusions and outline prospects for improvements to the models.

\section{Data and brief description of the spectra}

The data discussed in this paper were obtained during two observations with
XMM-Newton, as described in \citet{Ness11}. The first observation started 
at day 39.86, measured from the initial optical discovery on 2008 April 10.73 UT,
and lasted  39280\,s. During this first observation the RGS countrate varied strongly:
from about 15 cts/s during the first 7500\,s it dropped to a minimum near
3 cts/s lasting some 6000\,s, and then rose to a maximum of about 18 cts/s
for the final 16000\,s of the observation. We follow \citet{Ness11} in defining
three spectra for this first observation, Spectrum\,A.1 obtained
during the initial plateau;  Spectrum\,A.2 during the minimum flux; and 
Spectrum\,A.3 during the final high-flux plateau,  \citep[see Fig.\,3 of][]{Ness11}.
The second observation started on day 49.62, and showed a countrate which
varied only mildly around 3 cts/s during the 31700\,s observation. All data from
this observation are collected in the fourth spectrum, Spectrum B.

We analyze the 7-38\,\AA\ (1.77-0.33\,keV) first order spectra of the RGS detector,
using SPEX version 2.02.03 \citep{kaastraspex}.

\begin{figure*}
    \centering
 \vspace{-0.5cm}
\includegraphics[width=0.64\textwidth, angle=-90]{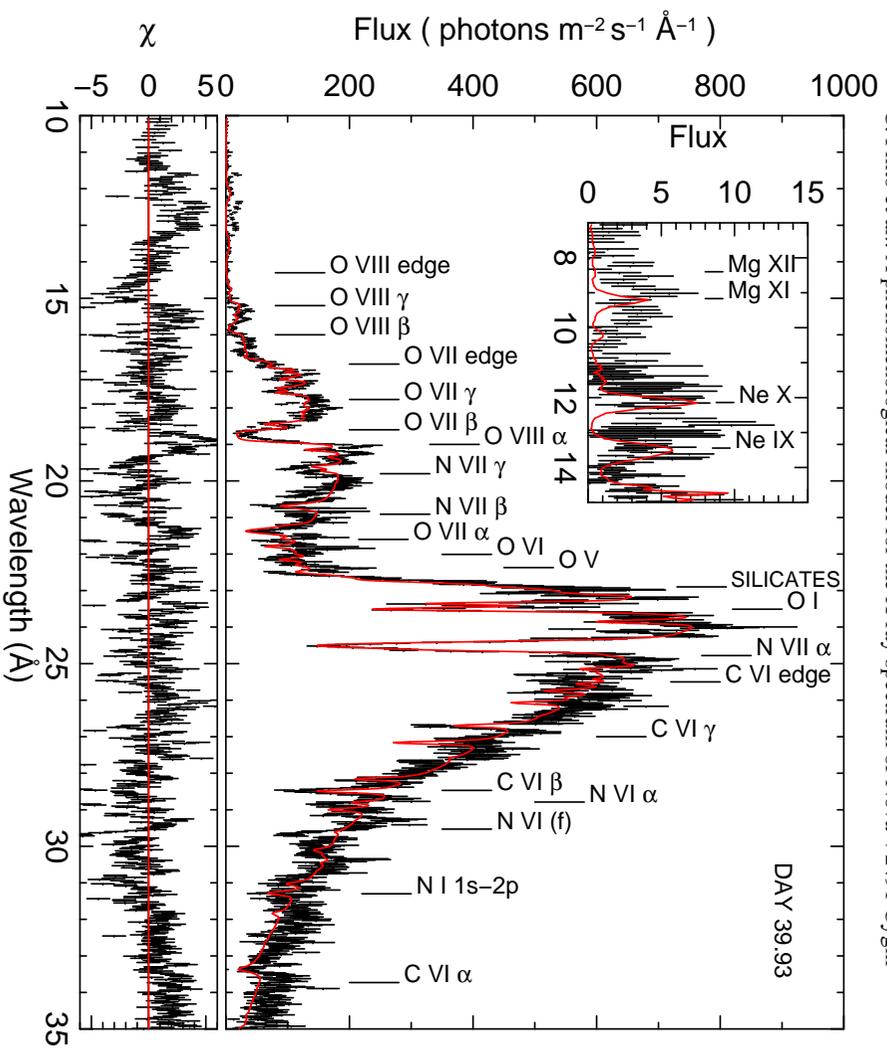}
\caption{Best fit to the first observation A.1 (see also Table \ref{table:rgs_fit_colors}). The rest-frame wavelengths of all the main transitions are displayed.}
    \label{fig:rgs_obs1}
   \end{figure*}

\begin{figure*}
    \centering
 \vspace{-0.5cm}
\includegraphics[width=0.64\textwidth, angle=-90]{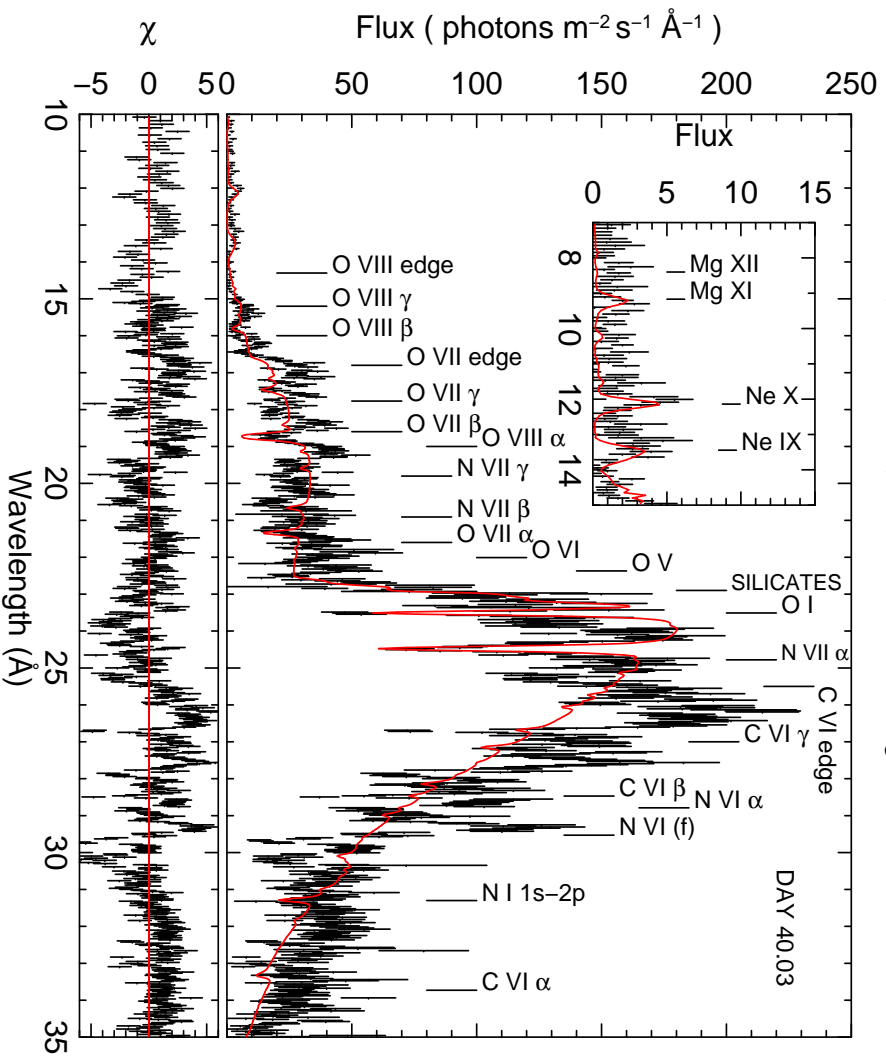}
\caption{Best fit to the second observation A.2.}
    \label{fig:rgs_obs2}
   \end{figure*}

\begin{figure*}
    \centering
 \vspace{-0.2cm}
\includegraphics[width=0.62\textwidth, angle=-90]{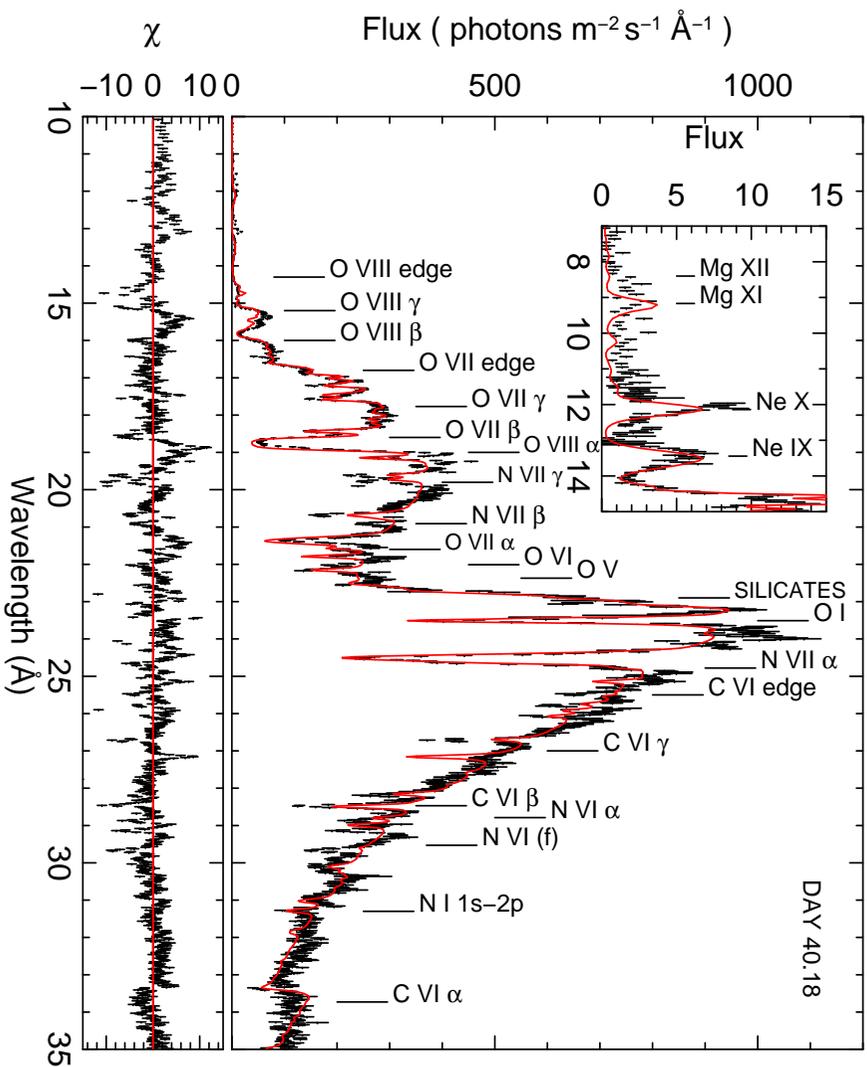}
\caption{Best fit to the third observation A.3.}
    \label{fig:rgs_obs3}
   \end{figure*}

\begin{figure*}
    \centering
 \vspace{-0.5cm}
\includegraphics[width=0.64\textwidth, angle=-90]{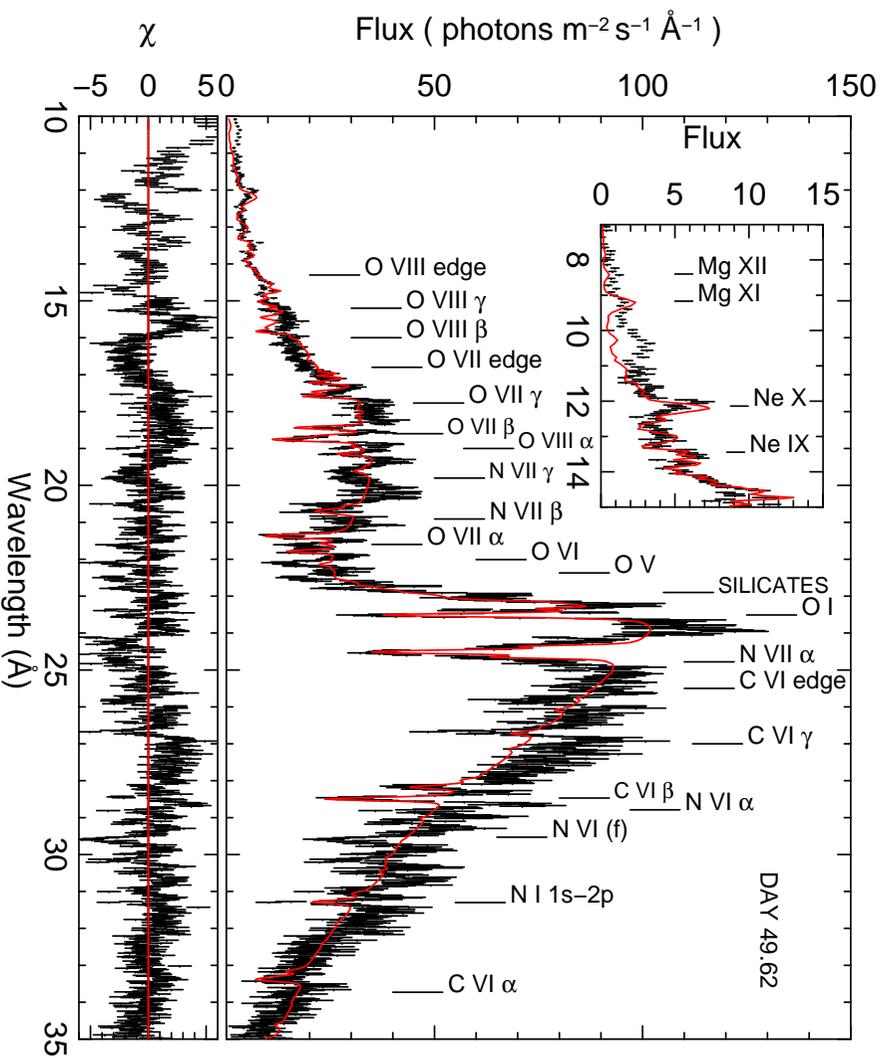}
\caption{Best fit to the fourth observation B.}
    \label{fig:rgs_obs4}
   \end{figure*}

We show the four spectra derived from the XMM-Newton RGS observations in Figs.
\ref{fig:rgs_obs1}-\ref{fig:rgs_obs4}  \citep[see also Figs.\ 6,7 of][]{Ness11}.
The most prominent features are the neutral oxygen absorption edge
near 23\,\AA\ with the corresponding 1s-2p absorption line at 23.5\,\AA.
The neutral nitrogen 1s-2p absorption line near 31.2\,\AA\ is also
seen, but the neutral iron L$_2$ and L$_3$ edges that one expects
near 17.15 and 17.45\,\AA\ are weak. Ness et al.\ (2011) identify 1s-3p
and 1s-4p lines of \ion{O}{vii}  near these wavelengths.
Indeed, absorption lines from \ion{O}{v} to \ion{O}{viii} are detected, blueshifted with
respect to the rest frame wavelength; examples include \ion{O}{viii} near
19\,\AA, \ion{O}{vii} at 21.5\,\AA, and \ion{N}{vii} at 24.5\,\AA. The blueshift
indicates that the absorbing matter is expanding, and the wide range
of ionization levels indicates that several expansion shells are
involved.


Lines of highly ionized neon and oxygen are also detected at the rest
frame wavelengths; examples are \ion{Ne}{x} at 12.1\,\AA, \ion{Ne}{ix}
at 13.4\,\AA, \ion{O}{viii} at 19.0\,\AA, and \ion{O}{vii} at
21.6\,\AA. A strong feature is also seen near 29.3\,\AA, close to the
forbidden \ion{N}{vi} line at 29.5\,\AA.

\section{The basic model}
  \label{sect:the_basic_model}

As a first attempt to describe the spectrum $F_\lambda$ we use a multi-component model
which may be written symbolically as
\begin{equation}
 F_\lambda =\left(F_{\lambda,\mathrm{BB}}\prod_{i=1}^3X_i
 + F_{\lambda,\mathrm{C}}\right) \cdot X_g \cdot X_d
\end{equation}

The components in this model and their parameters are
\begin{itemize}
\item the blackbody flux $F_{\lambda,\mathrm{BB}}(R,T_\mathrm{eff})$ characterized by
the radius $R$ and effective temperature $T_\mathrm{eff}$,
\item three photo-ionized absorbers
  $X_i(N_{i,O},\xi_i,v_i,\sigma_i,(A/O)_i)$, each characterized by the
  column density for oxygen $N_{i,O}$, ionization parameter $\xi_i$,
  outflow velocity $v_i$, velocity width $\sigma_i$, and the
  abundances of various elements with respect to oxygen $(A/O)_i$.
  {The number of absorbers is in principle free: we add absorbers until the fit does not significantly
  improve anymore (as indicated by the $\Delta\,\chi^2$; see Table~\ref{table:rgs_fit_colors}). 
  In this case, this leads to at least three absorbers.}
\item the flux from a collisionally ionized plasma
  $F_{\lambda,\mathrm{C}}(\mathrm{EM},T,\sigma_\mathrm{C},(A_{Ne})_C,(Mg/Ne)_C)$,
  characterized by the emission measure $\mathrm{EM}\equiv\int
  n_en_XdV$, the temperature $T$, the turbulent velocity
  $\sigma_\mathrm{C}$, the neon abundance $(A_{Ne})_C$, and
  the abundance ratio of magnesium to neon $(Mg/Ne)_C$,
\item {the absorption by interstellar or circumstellar gas}
$X_g(N_H,T_g,A_g)$,
characterized by the hydrogen column $N_H$, the temperature $T_g$, 
and the abundances for each element $A_g$, and
\item {the absorption by interstellar dust} $X_d(N_{d,O})$, characterized by the
column of the oxygen in the dust $N_{d,O}$.
\end{itemize}

To correctly describe the interstellar absorption $X_g$ with high
spectral resolution, we use the SPEX model {\em hot}, putting its
temperature $T_g$ at a low value (near 1\,eV), suitable for the
(nearly) neutral component of the ISM.  This model describes the
transmission through a layer of collisionally ionized plasma.  We give
abundances of the ISM normalized on proto-solar values; here and in the
following, when we refer to proto-solar values we use those given
by \citet{Lodders09}.

The detailed form of the \ion{Fe} L edge near 17.2-17.7\,\AA\ and the \ion{O} K
edge near 22.7-23.2\,\AA\ may be affected by the presence of these
elements in molecules in dust \citep{paerels}. The absorption
$X_d$ by such molecules is described in SPEX with {\em amol} \citep{PintoISM2010}.
In the spectra of V2491\,Cyg, the iron feature is
weak, and thus we only constrain oxygen compounds in the interstellar
dust.

\begin{figure}
     \centering 
     \subfigure{
     \includegraphics[angle=90, width=0.5\textwidth]{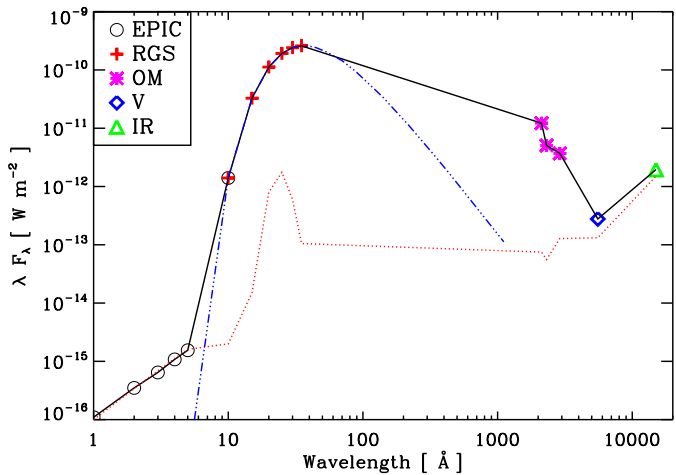}}
 \vspace{-0.5cm}
     \caption{The unabsorbed SED used for V2491 Cyg (solid line and data points).
              We also show the unabsorbed blackbody emission (blue dashed-dotted line) and the observed absorbed fluxes 
              (red dotted line).}
     \label{fig:sed}
    \end{figure}

A highly ionized absorber $X_i$ is modeled in SPEX with the {\em xabs}
model, which gives the transmission of a slab of photo-ionized
material.  Before this model is applied, one must compute the
ionization levels, and for this we use version C08.00 of Cloudy \citep[for a
description of an earlier version, see][]{Ferland98}. Cloudy
computes the ionization balance of a gas for the ionization parameter
$\xi \equiv L/(nr^2)$, a measure of the number of photons per
particle (for a light source with luminosity $L$ at distance $r$ of
the gas with number density $n$), and a given spectral energy
distribution (SED).
The form of the energy distribution is important, because a soft
irradiating spectrum leads to a different ionization equilibrium than
a hard irradiating spectrum with the same luminosity.
The energy distribution that we use for V2491\,Cyg
is shown in Fig.\,\ref{fig:sed}.
It has been computed through all the available archival data from IR to X-ray energies: IR fluxes from \citet{Naik09}, optical from \citet{Hachisu2009}, UV and soft X-ray from \citet{Ness11}, and hard X-ray from \citet{Page2010}. We have taken their fluxes and models in each wavelength band and corrected for the Galactic interstellar extinction by assuming a $N_{\rm H} = 2.5\times10^{25}$m$^{-2}$, which represents an average between our fits and those found in the literature. {The total luminosity of this SED is $1.15\times10^{33}$\,W, of which the X-rays, roughly described as a black body with the temperature of 110\,eV (blue dash-dotted line in Fig.\,\ref{fig:sed}), contributes $6.23\times10^{32}$\,W. }

The abundances of the highly ionized absorber obviously also are
important for the calculation of the ionization structure. In the
computation of the ionization levels we use proto-solar abundances.
In principle, once we have determined the abundances of the highly
ionized absorbers from our spectral fit, the computation of the
ionization structure should be iterated for these new abundances.  In
this pilot study, however, we have not performed this iteration.

To limit the number of free parameters, we limit the number of
independent abundances in the basic model as follows.  The flux
$F_{\lambda,C}$ of collisionally ionized plasma dominates only at
wavelengths $\lambda<15$\,\AA\ in the spectra of V2491\,Cyg and at
these wavelengths only magnesium and neon contribute to the emission.
We determine the Mg and Ne abundances from this part of the spectrum,
and assume that the abundances thus determined for the collisionally
ionized plasma are also valid for the highly ionized absorbers $X_i$.
Conversely, we determine the column of O, and the number ratios of C,
N, Si, S, Ar, Ca, Fe with respect to oxygen from the absorption lines
of the highly ionized absorbers $X_i$, assuming that these ratios are
the same for all three highly ionized absorbing shells.  We then
assume that the ratios found for the highly ionized absorbing shells
also apply to the collisionally ionized plasma.

The reason for listing abundances with respect to oxygen is that these
are much better constrained by the observed spectrum than abundances
with respect to hydrogen, because hydrogen only contributes to the
X-ray spectrum as a source of free electrons. The ionized hydrogen of the hot nova shell
only affects the spectral normalization but not the shape in the RGS spectra.

\subsection{Application to Spectrum\,A.3}

We apply the basic model first to the spectrum with the highest
signal-to-noise, Spectrum\,A.3.
In order to allow the use of $\chi^2$ statistics, we bin the RGS data
in bins of about 1/3 of the spectral resolution of $\sim$0.065\AA\
of the first order RGS spectrum. To ensure a minimum of 10 counts/bin
we in addition rebin the raw data at the shortest wavelengths. This leads
to rebinning by factors 10, 5 and 2 in the ranges 7-11\,\AA,
11-15\,\AA,  and 15-38\,\AA, respectively.

In a first step we fit a preliminary model, in which we ignore the 
highly ionized absorption shells $X_i$ and the dust absorption
lines $X_d$, This provides us with initial estimates for the
blackbody and collisionally ionized plasma, which we then enter
as starting values in the full model fitting.
The value that we derive for the temperature of the collisionally
ionized plasma is determined mostly from the spectrum at
$\lambda<15$\,\AA, and more specifically from the
ratio between the \ion{Ne}{ix} and \ion{Ne}{x} 1s-2p lines in this region.
For this reason we fix the plasma temperature $T$ 
in the full model fitting at the value found from the preliminary fit.

The resulting values of the parameters for the full model are listed in
Table\,\ref{table:rgs_fit_colors}, and the resulting model spectrum is
shown in Fig.\,\ref{fig:rgs_obs3}.

The radius of the blackbody emitter, 17\,600\,km, is several times 
the radius of a white dwarf (which decreases from  $\simeq$9000\,km
for a 0.6\,$M_\odot$ white dwarf to $\simeq$3000\,km at
1.3\,$M_\odot$; see Eq.(27) of \citet{Nauenberg1972}, and \citet{Provencal2002}).
This implies that the photospheric surface lies
in the expanding layers. With the temperature of $1.4\times10^6$\,K,
the blackbody has a bolometric luminosity $L_{BB}=8.5\times10^{32}$\,W.

To describe the high-ionization absorption lines, we require three
distinct $X_i$ components. An outflow with constant velocity and
constant mass flux $\dot M$,  has $n r^2\propto \dot M/v$ and
therefore $\xi$ is constant with radius. If the velocity increases
with radius, as in a homologous expansion, we expect $\xi$ to increase
with radius. This agrees with our finding that the shell with the
highest ionization parameter $\xi_i$ has the highest
velocity. However, our finding that the shell with the intermediate
velocity has the lowest $\xi_i$ is puzzling.  We think that this
result may be due to systematic errors in the velocities of weak
shells 2 and 3, caused by the difficulty in separating the velocity
dispersion and shift in the absorption lines of these relatively weak
shells (see also Sect.\,\ref{subsec:uncertainties}).

The lines of \ion{C}{vi}, \ion{Ar}{xvi} and \ion{S}{xiv} are weak relative to the oxygen
lines, and this results in number ratios C/O, Ar/O and S/O
much lower than solar. For Ca we can only give an upper limit
to the Ca/O ratio. 
We estimate the hydrogen column by assuming a proto-solar O/H ratio.

The relative strengths of the \ion{O}{vii}\,$\gamma$ plus \ion{Fe}{i}\,L edge at
17.2-17.7\,\AA\ and the oxygen features at 22.7-23.2\,\AA\ indicate an
Fe/O ratio lower than proto-solar. This suggests that Fe may be
depleted by dust. Indeed, we prefer to describe the broad absorption
feature at 22.7-23.3\,\AA\ with a dusty-molecule component of silicates
plus a gaseous component of \ion{O}{ii}, with about 10\%\ of the total
oxygen in each of these two components.  The alternative explanation
with low-ionization \ion{O}{ii-iv} in gaseous form only is less appealing as
it requires a very high velocity dispersion.

\subsubsection{Deviations}
\label{subsec:failures}

There are some features that limit the quality of the fit. Most of them show broad emission-like features and are much more relevant in the low continuum spectra. Indeed, in the high flux spectra these features are well hidden by both strong continuum and absorption lines. The strongest lines are at 18.4\,{\AA}, 19.0\,{\AA}, 19.2\,{\AA}, 21.5\,{\AA}, 21.9\,{\AA}, 26.5\,{\AA}, 27.5\,{\AA} and 29.35\,{\AA}. Most of the emission lines are blue-shifted less than the absorption line of the same transition, thus the emission should originate from a broader region of the shell. This would explain the whole line shape similar to a P$\,$Cygni profile. There are also a few absorption lines not reproduced by the model at 19.8\,{\AA}, 25.8\,{\AA}, 29.1\,{\AA} and 29.7\,{\AA}, which have already been reported by \citet{Ness11} as unidentified lines. These lines could represent an additional slower ($-2\,000$ km s$^{-1}$) expanding layer as their projected wavelengths are consistent with \ion{O}{v}, \ion{Ca}{xiii}, \ion{Ar}{xiii} and \ion{Fe}{xvi}/\ion{Al}{xi} transitions. However, as mentioned above, these absorption lines are so much weaker than the strong ones, that the inclusion of an additional layer does not provide a significant improvement to the fit.

\subsubsection{Systematic errors} 
\label{subsec:uncertainties}

In Table \ref{table:rgs_fit_colors}, only the relatively small statistical parameter errors are given which are likely much smaller than the combined statistical plus systematic uncertainties. The statistical errors are a proxy for the quality data, while systematic uncertainties are the main limitation for the accuracy of our results. A few parameters are highly model-dependent or have a degenerate value. Both the velocity dispersion and the blue-shift of the shell layers have this problem. The complexity of the absorption lines plus the presence of weak emission lines complicate the modeling as the different components could mix with each other. This can happen to the components that refer to the weak unresolved emission lines (layer 2 and 3). Thus, the conclusion of different expansion velocities in two low-ionization layers is less certain than, e.g., the fact that the highest-ionization layer has a higher velocity shift and broadening.
%

\begin{table*}
\caption{RGS best fit parameters to the four observations (see also Fig. \ref{fig:rgs_obs1}, \ref{fig:rgs_obs2}, \ref{fig:rgs_obs3}, and \ref{fig:rgs_obs4}).} 
\begin{center}
\small\addtolength{\tabcolsep}{+10pt}
\scalebox{1}{%
\begin{tabular}{|l|l|l|l|l|l|l|}
\hline
 Component            & Parameter                    &  A.1     &  A.2     &  A.3     &  B     \\
\hline
\hline
\multirow{3}{*}{Blackbody} & R ($10^{7}$ m)             &   $2.1\pm0.2$  & $1.0\pm0.1$    &  $1.76\pm0.03$ &  $0.19\pm0.05$\\
                       & $kT_{\rm eff}$ (eV) \index{\footnote{}}  &   $91\pm1$     & $81\pm1 $      &  $121\pm2$     &  $  95\pm1 $ \\
                   & ${\rm L}_{\rm RGS}$ ($10^{32}$ W)  &  $2.20       $ &  $0.23       $ &  $5.70       $  &  $0.03       $ \\
                   & ${\rm L}_{\rm BOL}$ ($10^{32}$ W)  &  $4.39       $ &  $0.57       $ &  $8.49       $  &  $0.04       $ \\
\hline
\multirow{7}{*}{CIE} & $n$e $n$H V (10$^{63}$ m$^{-3}$) &  $5.6\pm0.7$   &  $7.1\pm0.7$   &   $9.6\pm0.4$   &  $5.2\pm0.4$\\
                   & $kT$ (keV) $^{(a)}$                &  $0.39\pm0.05$ &  $0.39\pm0.05$ &  $0.37\pm0.05$  &  $0.71\pm0.05$ \\
                   & ${\rm L}_{\rm RGS}$ ($10^{28}$ W)  &  $0.79       $ &  $1.05       $ &  $1.40       $  &  $0.94       $ \\
                   & ${\rm L}_{\rm BOL}$ ($10^{28}$ W)  &  $1.24       $ &  $1.65       $ &  $2.24       $  &  $1.45       $ \\
                     & $\sigma_C$ (km s$^{-1}$)         &  $\equiv$ 3000 &  $\equiv$ 3000 &   $3000\pm300$  &  $\equiv$ 3000 \\
                     & Mg / Ne                          &  $\equiv  1.2$ &  $\equiv  1.2$ &   $1.2\pm0.2$   &  $\equiv  1.2$ \\
                        & $\Delta \chi^2$ / dof         & $ 85/2 $       & $   136/2  $   & $ 709/4  $      & $ 255/2 $      \\
\hline
\hline
\multirow{5}{*}{Abundances} &   C / O            & $\equiv 0.13 $ &  $\equiv 0.13 $ &  $ 0.13 \pm 0.01  $ &   $\equiv 0.13 $ \\
                            &   N / O            & $\equiv 2.41 $ &  $\equiv 2.41 $ &  $ 2.41  \pm 0.01 $ &   $\equiv 2.41 $ \\
                            &   Si / O           & $\equiv 0.015$ &  $\equiv 0.015$ &  $ 0.015 \pm 0.005$ &   $\equiv 0.015$ \\
     in the shell           &   S / O            & $\equiv 0.11 $ &  $\equiv 0.11 $ &  $ 0.11 \pm 0.01$   &   $\equiv 0.11 $ \\
                            &   Ar / O           & $\equiv 0.20 $ &  $\equiv 0.20 $ &  $ 0.20 \pm 0.01$   &   $\equiv 0.20 $ \\
                            &   Ca / O  $^{(b)}$ & $\equiv  0.01$ &  $\equiv  0.01$ &  $ \lesssim   0.01$ &   $\equiv  0.01$ \\
                            &   Fe / O           & $\equiv 0.47 $ &  $\equiv 0.47 $ &  $ 0.47 \pm 0.01$   &   $\equiv 0.47 $ \\
\hline
\hline
\multirow{6}{*}{Layer 1}& H Col. (10$^{28}$ m$^{-2}$)  & $ 0.73\pm0.02$ & $ 2.13\pm0.01$ & $ 0.48\pm0.01$ & $ 1.22\pm0.07$ \\
                        & O Col. (10$^{25}$ m$^{-2}$)  & $ 0.44\pm0.01$ & $ 1.29\pm0.01$ & $ 0.29\pm0.01$ & $ 0.74\pm0.04$ \\
                        &  Log $\xi$ (10$^{-9}$ Wm)    & $ \gtrsim 5.0$ & $ 4.25\pm0.02$ & $ \gtrsim 4.9$ & $ 4.38\pm0.06$ \\
                        & $\sigma_V$     (km s$^{-1}$) & $ 1230\pm20  $ & $  225\pm35  $ & $ 1470\pm10  $ & $   55\pm20 $  \\
                        & $v$     (km s$^{-1}$) & $-3730\pm30  $ & $-3360\pm70  $ & $-3620\pm20  $ & $-4560\pm130 $ \\
                        & $\Delta \chi^2$ / dof        & $ 526/4 $      & $ 609/4  $     & $ 4083/11  $   & $ 449/4 $      \\
\hline
\multirow{6}{*}{Layer 2}& H Col. (10$^{28}$ m$^{-2}$)  & $  2.0\pm0.2 $ & $  0.1\pm0.05$ & $ 4.15\pm0.02$ & $0.013\pm0.002$\\
                        & O Col. (10$^{25}$ m$^{-2}$)  & $  1.2\pm0.1 $ & $ 0.06\pm0.03$ & $ 2.51\pm0.01$ & $0.008\pm0.001$ \\
                        &  Log $\xi$ (10$^{-9}$ Wm)    & $ 3.61\pm0.01$ & $ 2.50\pm0.05$ & $ 3.76\pm0.01$ & $ 2.18\pm0.03$ \\
                        & $\sigma_V$     (km s$^{-1}$) & $   10\pm5   $ & $   10\pm5   $ & $   20\pm5   $ & $  160\pm10  $ \\
                        & $v$     (km s$^{-1}$) & $-2790\pm20  $ & $-3260\pm20  $ & $-2810\pm10  $ & $-3080\pm40  $ \\
                        & $\Delta \chi^2$ / dof        & $ 526/4 $      & $ 62/4  $      & $ 2951/11  $   & $ 373/4 $      \\
\hline
\multirow{6}{*}{Layer 3}& H Col. (10$^{25}$ m$^{-2}$)  & $  8.1\pm0.2$  & $ 0.5\pm0.1$     & $  8.1\pm0.2$   & $  2.6\pm0.2 $ \\
                        & O Col. (10$^{22}$ m$^{-2}$)  & $  4.9\pm0.1$  & $ 0.30\pm0.06$   & $  4.9\pm0.1$   & $  1.6\pm0.1 $ \\
                        &  Log $\xi$ (10$^{-9}$ Wm)    & $ 1.40\pm0.05$ & $\lesssim 0.01$  & $ 1.36\pm0.01$  & $ 1.18\pm0.05$ \\
                        & $\sigma_V$     (km s$^{-1}$) & $  235\pm10 $  & $   70\pm20  $   & $  200\pm10  $  & $  180\pm20  $ \\
                        & $v$     (km s$^{-1}$) & $-3400\pm30 $  & $\gtrsim -3040 $ & $-3340\pm20 $   & $-3300\pm50  $ \\
                        & $\Delta \chi^2$ / dof        & $ 448/4 $      & $   25/4  $      & $  781/11  $    & $ 399/4 $      \\
\hline
\hline
\multirow{4}{*}{Cold gas}& Col. (10$^{25}$ m$^{-2}$)   &  $2.85\pm0.01$ &  $2.81\pm0.01$ &  $2.24\pm0.01$ &  $1.97\pm0.02$ \\
                        & $kT$ (eV)         &  $1.13\pm0.01$ &  $1.21\pm0.01$ &  $1.04\pm0.01$ &  $0.99\pm0.02$ \\
                        & N / H                        &  $\equiv 2.14$ &  $\equiv 2.14$ &  $2.14\pm0.02$ &  $\equiv 2.14$ \\
                        & O / H                        &  $\equiv 2.71$ &  $\equiv 2.71$ &  $2.71\pm0.01$ &  $\equiv 2.71$ \\
                        & Fe / H                       &  $\equiv 1.19$ &  $\equiv 1.19$ &  $1.19\pm0.03$ &  $\equiv 1.19$ \\
\hline
\multirow{2}{*}{Dust}   &\ion{O}{i} (10$^{21}$ m$^{-2}$)&  $3.8\pm0.1$  &  $6.3\pm0.1$   &  $3.2\pm0.1$   &  $5.3\pm0.2$   \\
                        & $\Delta \chi^2$ / dof        & $ 497/1 $      & $   356/1  $   & $ 1300/1  $    & $ 948/1 $      \\
\hline
\hline
Statistics                   &    $\chi^2$ / dof       &  3538/1474     &  3980/1474     & 10380/1462     &  3390/1474     \\
 \hline
\end{tabular}}
\label{table:rgs_fit_colors}
\end{center}
$^{(a)}$ The CIE temperature is kept frozen to the value estimated with a local fit to the $7-15$ {\AA} spectral range. $^{(b)}$ We can only provide upper-limit for Ca relative abundance. $^{(c)}$ All the errors shown are statistical, systematic effects are not considered here (for further information see Sect \ref{subsec:uncertainties}).
\end{table*}

\subsection{Application to Spectra A.1, A.2 and B}

In fitting the other X-ray spectra of V2491\,Cyg, which have rather
lower signal to noise, we fix the velocity dispersion of the
collisionally ionized plasma, as well as the (relative) abundances of
this plasma, of the highly ionized absorbers $X_i$, and of the cold
absorption component $X_g$ to the values determined for Spectrum\,A.3.
The resulting values of the other parameters are collected in
Table\,\ref{table:rgs_fit_colors}, and the spectral fits are shown in
Figs.\,\ref{fig:rgs_obs1}, \ref{fig:rgs_obs2} and
\ref{fig:rgs_obs4}, respectively.

The blackbody radius of Spectrum\,A.1 is similar to that of
Spectrum\,A.3, but the temperature is lower by a quarter, and as a
result the bolometric luminosity is lower by 50\%; within the RGS band
the blackbody flux is lower by 60\%.  For the
collisionally ionized plasma, the temperature is the same in the
spectrum Spectrum\,A.1 as in Spectrum\,A.3, but the emission
measure, and therefore also the luminosity, is lower by 45\%\ in the
former.  As regards the highly ionized absorption shells, all
parameters are similar in the Spectra\,A.1 and A.3, with
the exception of the column of the shell with the highest ionization,
1.5 times bigger in A.1, and the column of the shell
with the middle ionization, which in  Spectrum\,A.1 was 0.5
times that of A.3. Finally, the columns of the cold absorbers
$X_g$ and $X_d$ both are markedly higher, by factors 1.3 and 1.2
respectively, in Spectrum\,A.1 with respect to A.3.

The blackbody radius of Spectrum A.2 is only 10\,000\,km, and
thus closer to the white dwarf surface. One might expect a higher
temperature there, but the observation shows a lower temperature, 
ten percent lower even than the low temperature of Spectrum\,A.1. The
luminosity of the blackbody accordingly is just 10\%\ in Spectrum\,A.2
of that in Spectrum\,A.1.  The emission measure of the
collisionally ionized plasma in Spectrum\,A.2 is intermediate between those
before in Spectrum\,A.1 and after in Spectrum\,A.3, at unchanged
temperature. The ionization parameter of all three high-ionization
absorption shells is lower in Spectrum A.2. The column of the
shell with the highest ionization has increased by a factor of
almost three between Spectrum\,A.1 and Spectrum\,A.2; in contrast the columns
of the other two shells have plunged.  The column of the cold
absorbing gas is the same in Spectrum A.2 and Spectrum\,A.1, but the 
column of the dust much higher in Spectrum A.2.

The large changes in the spectral components well within the eleven
hours of the first XMM-Newton observation are striking; we will
investigate these some more in Sects.\,\ref{s:variants} and \ref{s:disc}.

Almost ten days later, during the second observation of V2491\,Cyg
with XMM-Newton, the radius of the blackbody component has shrunk to
1900\,km. This remarkably low value would imply that only part of the
surface of the white dwarf has this high temperature; or alternatively
that the white dwarf is more massive than 1.3\,$M_\odot$. The
luminosity of the blackbody component has declined to
$L_\mathrm{BB}=3.8\times10^{30}$\,W. The emission measure of the collisionally ionized plasma
has decreased, but its temperature has increased, and as a result its
luminosity now is $L_\mathrm{CIE}=1.45\times10^{28}$\,W, to a large
extent within the RGS band (see Table\,\ref{table:rgs_fit_colors}).

The ionization levels of the three highly ionized absorbers have
dropped, as have the columns of shells 2 and 3, between Spectra\,A.1
or A.3 and Spectrum\,B; the column of the highest ionization shell,
shell 1, has increased.  The column of the cold gas absorber has
dropped by some 10\%, but the column of the dust component has
increased by a factor 1.7. In general, the relative changes of the
parameters of Spectrum\,B with respect to those of Spectra A.1 and A.3
are similar qualitatively, if not quantitatively, as the changes of the
spectrum with the lowest countrate during the first observation, i.e.\
Spectrum A.2.

\section{Variant models}\label{s:variants}

To test the robustness of some of the results from the basic models,
we have fitted some variant models. The topics that we wish to
investigate in particular are the nature of the hot absorption shells
and the cause of the changes during the first observation.

\subsection{The hot absorption shells and the bolometric luminosity}

In the basic model the hot ionized absorber consists of three
shells, all with the same abundances, but with distinct ionization
parameters $\xi_i$.

\begin{figure}
     \centering 
     \subfigure{
     \includegraphics[bb=53 53 526 683, angle=90, width=0.5\textwidth]{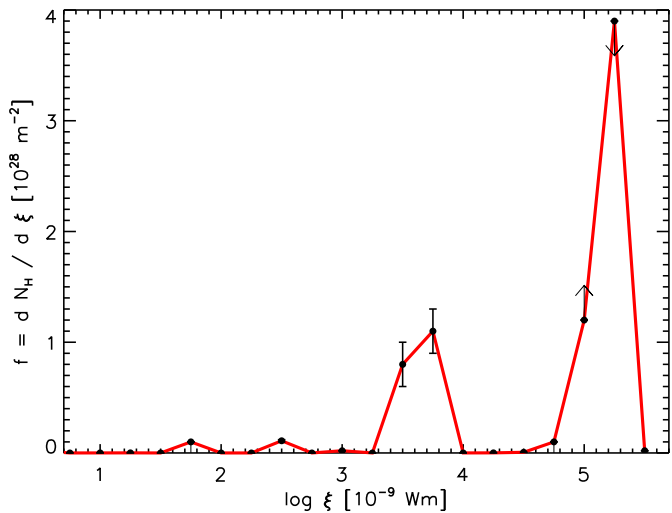}}
 \vspace{-0.5cm}
     \caption{Absorption measure distribution (see also Table~\ref{tab:warm}).}
     \label{fig:warm}
    \end{figure}

As an alternative we have fitted a continuous absorption measure distribution (AMD),
using the SPEX model {\sl warm}, to Spectrum\,A.3 {(see Table~\ref{tab:warm} and Fig.~\ref{fig:warm})}. 
We find that an initially continuous distribution in the course of the fitting
procedure collapses to discrete components, compatible
with the findings of our basic model. {The model confirms the presence of at least two discrete ionization ranges with discrete velocities,  represented by $\log\,\xi=3.50-3.75$ and $\log\,\xi\geq4.75$, which are consistent with what we have found 
with separate {\sl xabs} components (see Table~\ref{table:rgs_fit_colors}).} The {\em
  warm} model uses the same velocity for the whole range of $\xi$, but
the observed spectrum shows a relation between ionization and
velocity, and this causes the {\em warm} model fit to be decidedly
worse than the basic model {($\chi^2$/dof\,=\,17300\,/\,1480)}.

\begin{table}
\caption{Absorption measure distribution.}
\begin{tabular}{llllll}
\hline
 $\xi^{b}$ & $f^{a,b,c}$ & $\xi^{b}$ & $f^{a,b,c}$ & $\xi^{b}$ & $f^{a,b,c}$  \\
\hline
0.50 &  $\leq 2\times10^{-2}$  & 2.50 & $0.11\pm0.01$          &  4.50 & $\leq 6\times10^{-3}$\\
0.75 &  $\leq 5\times10^{-6}$  & 2.75 & $\leq 1\times10^{-15}$ &  4.75 & $\leq 0.1$ \\
1.00 &  $\leq 5\times10^{-4}$  & 3.00 & $\leq 2\times10^{-2}$  &  5.00 & $\geq 1.2$ \\
1.25 &  $\leq 1\times10^{-3}$  & 3.25 & $\leq 4\times10^{-8}$  &  5.25 & $\leq 3.9$      \\
1.50 &  $\leq 1\times10^{-11}$ & 3.50 & $0.8\pm0.2$            &  5.50 & $\leq 0.02$      \\
1.75 &  $\leq 1\times10^{-2}$  & 3.75 & $1.1\pm0.2$            &          \\
2.00 &  $\leq 1\times10^{-10}$ & 4.00 & $\leq 4\times10^{-4}$  &          \\
2.25 &  $\leq 1\times10^{-15}$ & 4.25 & $\leq 1\times10^{-3}$  &          \\
\hline
\end{tabular}
\vspace{0.1cm}

$^{a}$ $f = {\rm d}\,N_{\rm H} / {\rm d}\,\log\,\xi$  \\
$^{b}$ $\xi$ is in units of 10$^{-9}$ Wm and $N_{\rm H}$ in 10$^{28}$ m$^{-2}$\\
$^{c}$ The values refer to the {\sl warm} model fits to spectrum\,A.3. \label{tab:warm}
\vspace{-.1cm}
\end{table}
\vspace{-.1cm}

We have also fitted Spectrum\,A.3 with some models in which we fit the
abundances separately for each shell; this does not improve the fits
significantly.

{While the limited number of absorbers suggests the presence of discrete shells with
different velocities and a non-monotonous $\xi$ distribution, we
note that any number of absorbers is possible and therefore a continuous 
distribution. The two strong peaks in the absorption measure distribution suggests
the presence of two main absorbers. However, the high velocities make difficult to believe
that the shells might be stable throughout the first month after the outburst and argue in favor
of a continuous distribution.}

The temperatures of the hot absorbing shells imply that hydrogen is
fully ionized. This means that the hydrogen column of these shells is
badly constrained, and for this reason we give the abundances of the
various elements with respect to oxygen, fixing the hydrogen column to
the proto-solar hydrogen-to-oxygen ratio in the basic model.  The main
effect of the presence of hydrogen is that photons that initially move
from the central source in our direction are scattered into another
direction. In the fully spherically symmetric case, these photons are
replaced by photons moving initially in other directions that are
scattered into our direction by Thompson scattering. The slab model that we use for the
absorbing shell assumes that each photon from the central source that
is scattered within the shell does not reach the observer, and is {\em
  not} replaced by photons from initially other directions. In the
presence of hydrogen, the number of electrons in the shell is much
higher than in the absence of hydrogen, and a much higher number of
photons is scattered. For a given observed number of photons, the slab model
with hydrogen leads to a much higher number of source photons incident
on the shell than the model without hydrogen.

Whether this implementation of the model is valid in the case of
V2491\,Cyg is not clear. The rapid variations between spectra A.1, A.2
and A.3 may suggest that spherical symmetry does not apply.  If
spherical symmetry does apply, however, the number of photons should
not be corrected for scattering, i.e.\ the luminosity of the central
source will be much lower than found with the slab model.  To
investigate the magnitude of this effect we consider first a variant model
in which we set the hydrogen content of the three shells to zero, but keep the
helium. To good approximation, the electron scattering of the photons
from the central source does not depend on energy within the RGS
range, and thus the parameters of the fit are not affected, other than
that the luminosity of the central source is reduced. In particular,
the temperature of the blackbody is the same as in the basic
model. Thus we compute the luminosity for the blackbody emitter
for the model in which the photon number is not corrected for
electron scattering in the absorption shells by fixing all parameters
to the values of the basic model, and then recompute the
luminosity of the emitter for spectrum A.3. This luminosity, at
3.92\,10$^{31}$W, and with it the radius, at 3780\,km, are much
smaller than in the basic model.

This reasoning applies equally to a second variant model where we
set the hydrogen and helium content both to zero. For spectrum A.3 the
temperature again is the same as in the basic model, but the
luminosity and radius  drop to 2.14\,10$^{31}$W and 2790\,km,
respectively.
Table\,\ref{tab:variants} lists the variation of the parameters with
assumed hydrogen and/or helium content for the slab model, or
alternatively the variation with increasing spherical symmetry.

\begin{table}
\caption{Radii and luminosities of the central blackbody emitter.}
\begin{tabular}{l|rrr|rrr|r}
 & \multicolumn{3}{c}{$R$ (km)} &
 \multicolumn{3}{c}{$L_\mathrm{BB}$ (10$^{30}$\,W)} & $T_\mathrm{eff}$
 (ev)\\
\hline
 & basic & no H & no H\phantom{l} & basic & no H & no H\phantom{l} \\
 &            &         & no He &          &          & no He \\
\hline
A.1 & 21000 & 7000 & 5600 & 440  & 47   & 31   & 91\\
A.2 & 10000 & 4800 & 4100 & 57   & 13   & 9    & 81\\
A.3 & 17600 & 3780 & 2790 & 849  & 39.2 & 21.4 & 121 \\
B   & 1900  & 1500 & 1300 & 3.8  & 2.4  & 2.2  & 95 
\end{tabular}
\vspace{0.1cm}

The values refer to the fits to the four spectra, for the basic model, and for
  variant models in which the hydrogen content, or the hydrogen and
   helium content of the highly ionized absorption shells are set to
   zero. Note that the parameters for the basic model are correct only
   for a highly non-symmetric case; in the spherically symmetric case
   the correct values for the basic model are those listed above for
   the no H, no He case. \label{tab:variants}
\end{table}

{In computing the ionization equilibria with Cloudy we assume that the ultraviolet/optical flux is emitted by the central source. If instead it is emitted elsewhere, it may not contribute to the ionization of the shells. For this reason, we have calculated the 
ionization balance of the shells when the UV fluxes in the adopted SED were decreased by up to two orders 
of magnitude. The bolometric luminosity decreased by less than 5\%, while the newly-fitted ionization parameters of the shells 
are systematically lower by 15--25\%. The non-monotonous $\xi$ structure as seen in Fig.~\ref{fig:warm} is still preserved and 
the other main physical parameters like column densities and abundances are consistent within the errors. The overall scenario thus
is unaffected.
}

\subsection{The changes between Spectra A.2 and A.3}

The dramatic changes in brightness of the X-ray source during the
first observation with XMM-Newton are explained in the basic model by
a drop in the luminosity of the central blackbody emitter, combined
with changes in the column of the ionized absorption shells. The
luminosity drop of the central source is the result of a smaller
radius and a lower temperature (see Table\,\ref{tab:variants}). In the variant models
discussed in the previous section, where the hydrogen and helium
contents of the absorption shells are set to zero, the drop in
luminosity between A.3 and A.2 are much less pronounced, since in
these models the lower temperature of spectrum A.2 is partially offset
by a larger radius (Table\,\ref{tab:variants}). It is remarkable that
$R$ in the models with reduced hydrogen and helium contents
drops monotonically with time; in the basic model $R$ 
increases between spectrum A.2 and spectrum A.3.

To investigate whether it is possible to ascribe the change between
spectra A.2 and A.3 with a change in the blackbody and collisionally
ionized emitters only, we fitted spectrum A.2 with an alternative
model, in which all parameters related to the absorption components
are fixed to the values for Spectrum\,A.3, and only the parameters for
the blackbody and for the collisionally ionized plasma are allowed to
change.  The resulting fit is significantly worse ($\chi^2\simeq5000$,
as compared to the best basic fit $\chi^2\simeq 4000$).  The $\chi^2$
for this fit is dominated by the bad fit to the emission between
25-29\,\AA, and around 16\,\AA.

To investigate whether the lower luminosity may be explained by
increased absorption due to increased columns of the ionized 
shells only, we fit spectrum A.2 with a model in which the parameters
for the central emitters are fixed to the best values of the basic
model for spectrum A.3, and only the columns and ionization parameters
of the shells are allowed to vary. The best fit with this variant
model has $\chi^2\simeq5500$, significantly worse than the basic
fit for spectrum A.2.

To investigate whether the lower luminosity may be explained by
increased absorption due to an increased column of the neutral gas
and/or an increased column of the dust, we fit spectrum A.2 with a
model in which the columns of the neutral gas and of the dust are
fitted, but all other parameters are fixed to the best-fit parameters
of the basic model for spectrum A.3. This gives a bad fit, with
$\chi^2\simeq 8000$. A model in which the central emitter
is only partially absorbed by an increased neutral hydrogen column
does not fit the spectrum at all, due to the fact that the H\,I
absorption is strongest at $\lambda\gtap24$\,\AA, a part of the
spectrum that remains relatively bright. Thus, we can exclude absorption by cold clumpy material.

Finally, we have checked whether an addition of a radiative
recombination continuum (RRC) to the model of spectrum A.2 improves
the fit. The presence of this emission is suggested by emission
features near 16.8\,\AA\ and 18.4\,\AA. We take into account the
recombination of the most relevant ions of carbon, nitrogen and
oxygen. The addition of RRC to the model does not lead to a
significant improvement of the fit.

\subsection{{Derived mass loss rates and model consistency}} \label{s:massloss}


{Our fits to the spectra imply a mass loss rate:
\begin{equation}\label{e:mdot}
\dot M_w =\left({\Omega\over4\pi}\right)4\pi r^2 \mu m_Hf_c\,n_Hv = 
\left({\Omega\over4\pi}\right)4\pi \mu m_Hv {L\over\xi}
\end{equation} 
where $\mu$ is the mean atomic weight in units of the hydrogen mass
$m_H$,
$f_c$ is a factor allowing for clumpiness of the density distribution ($f_c\leq1$),
and where $\Omega=4\pi$ in the case of spherical symmetry.
The observed spectrum gives $\xi$ and $v$ for each shell, almost
independent of $L$.
With the parameters of the standard model fit to spectrum A.3, we
obtain the mass loss rates listed in Table~\ref{tab:massloss}.
Note that these values are valid for one time interval only -- i.e.\ the time of
maximum X-ray flux, and thus may overestimate the outflow-rate averaged
over the outburst.
From Eq.\,\ref{e:mdot} we note in addition that the derived value of $\dot M_w$ scales
with the luminosity $L$ of the central source, and thus will be smaller when the
layers are hydrogen- and/or helium-poor (see Table~3), and also when the
layers are spherically symmetric, in which case the {\tt xabs} model overestimates
the central luminosity.}

{The depths of the absorption lines, combined with the fitted
abundances, are converted to the hydrogen column density 
$N_H\equiv \int n_Hdr \simeq f_c\,n_H\Delta r$ where $\Delta R$ is a
typical length scale. 
Our model implies that this length scale is smaller than the distance to the
central source, $\Delta r\equiv\beta r<r$, and this may be used to
obtain an upper limit to the distance $r$:
\begin{equation}
{L/\xi\over N_H} = {n_Hr^2\over f_c\,n_H\beta r} \Rightarrow r = f_c\,\beta
{L/\xi\over N_H} \qquad (f_c\,,\beta<1)
\end{equation}
and from this a lower limit to the density
\begin{equation}
n_H = {L\over \xi r^2} = {\xi \over L}\left({N_H\over f_c\,\beta}\right)^2
\end{equation}
Note that the definition and computation of $\xi$ involves the
physical density (i.e.\ not the space-averaged density), and therefore
does not contain the clumpiness factor $f_c\,$. The mean atomic weight $\mu$ is 1.42 for proto-Solar H-He abundances and 3.58 for H-He abundances decreased by two orders of magnitude.}
\begin{table}
\caption{Derived mass loss rates and shell parameters.}
\begin{tabular}{l|ccc}
   & layer 1 & layer 2 & layer 3 \\
\hline
 & & & \\
$\dot M /(f_c\,\Omega\mu) (M_\odot/\mathrm{yr})$ & $1.3\times10^{-5}$ &
$1.4\times10^{-4}$  & $4.1\times10^{-2}$ \\
$r/(f_c\,\beta r_\mathrm{BB})$ & 127 & 202 & 2.6$\times10^7$ \\
$n_H\times(f_c\,\beta)^2$ & $2.2\times10^{18}$ & $1.2\times10^{19}$ & $1.8\times10^{11}$ \\
\end{tabular}
\label{tab:massloss}

\vspace*{0.35cm}
{The parameters listed are valid for the standard model for spectrum A.3. If the layers are depleted in hydrogen and/or helium the actual mass loss rate would be lower.}
\vspace*{-0.35cm}
\end{table}
{Another interesting parameter is the time scale at which the gas in the outflow
adjusts its ionization balance. This may be calculated from the ionization balance
and $\xi$-values of the fits. Through the SPEX tool \textit{rec\_time} we determine the \ion{O}{viii} recombination times for layers~1 and 2. They spread between a few (H-He rich slabs) to up-to-ten (H-He poor slabs) seconds, which means that the gas is responding instantaneously to the variations of the photo-ionizing continuum of the source and that the observed features are linked to the variability of the source.
}

\section{Discussion}\label{s:disc}

From our fits it is clear that the central blackbody emission
varies, during our first observation, rapidly, on a time scale
of hours.  The variations are likely exaggerated by the
slab model: in full spherical symmetry, and also in the
absence of hydrogen and helium in the ionized absorption
shells, the changes are still significant, but much less.
Forty days into the nova outburst, we may well be looking at shells
whose abundances are affected by nuclear reactions.
Indeed, the abundances that we derive for carbon, nitrogen and
oxygen indicate that such is the case.
This implies that the hydrogen and helium abundances
of absorption shells are much lower than solar, in comparison to the
elements just mentioned. Even in the absence of spherical
symmetry, therefore, the variation of the central source is
less pronounced than implied by our basic model, and probably
more in agreement with the numbers found for the models in which
the absorption shells contain no hydrogen and/or helium.

{The response time required for the ionization equilibrium of shells to respond to changes in the ionizing continuum is a few seconds, which means that the photo-ionization balance of the shells responds almost instantaneously to variations in the source flux and that
its luminosity might have indeed changed. The mass loss rate in our standard model for spectrum A.3 for the third shell in particular is  rather high
compared to the common limits $10^{-4}-10^{-3}M_\odot/\mathrm{yr}$ found for other novae 
\citep{Kovetz1987, Kato1989, Smith2006}. The derived mass loss rates 
for all three layers are lower in the models with reduced hydrogen
and/or helium content of the absorbing shells. }

{ As discussed in Section 4.1, the blue-shifted absorption lines are caused
 because photons from the central source moving initially in our direction
 are scattered into another direction. If the dimension of the shell is large
 with respect to the size of the white dwarf, and if the optical depth of the
 shell in the lines is not too large, this will lead to a P Cygni profile, as
 we see photons initially moving in other directions and then scattered into
 our direction in emission, both blue- and red-shifted. The emission and
 absorption should have equal strength. In the spectrum of V2491 Cyg
 this is clearly not the case. One explanation is that the shell is not large
 with respect to the white dwarf, so that a large fraction of the receding
 half  is occulted. Similarly, a high optical depth in the shell may cause
 photons scattered from the receding half of the shell to be reabsorbed before
 escaping in our direction. And finally, a large asymmetry in the shell also 
 may reduce the strength of the red-shifted emission.}

Carbon and sulfur are about an order of magnitude less abundant than
nitrogen and oxygen, and elements as silicon, argon and calcium are
much less abundant. This suggests that the white dwarf in V2491\,Cyg
is a O-Ne rather than a C-O white dwarf. The strength of the
\ion{Ne}{ix-x} lines as compared to the \ion{C}{v-vi} lines supports
this conclusion, as do observations in other
wavelength regions \citep{Lynch2008, Helton08, Naik09, Munari2010}.

The strongest feature in all X-ray spectra is the rest frame
\ion{O}{i} K-edge around 23.0\,\AA, which is well reproduced in the
fits together with the corresponding \ion{O}{i} 1s-2p absorption line
at 23.5\,\AA. Other features of the cold CSM/ISM are the \ion{Fe}{i}
L-edge around 17.5\,\AA\ and the \ion{N}{i} 1-2p absorption line at
31.2\,\AA, both well fitted by our models.  Neutral oxygen and
nitrogen are highly overabundant, whereas iron is slightly
overabundant, with respect to solar abundances.  This suggests that
part of the cold absorption originates in enriched circumstellar matter close to
V2491\,Cyg. The reduction of the column and the decrease in
temperature of the cold absorber between spectra A.1 and B suggests
that the CSM column is mostly from the current nova outburst, and still
expanding, albeit slowly. Page et al.\ (2010a) suggest that the
reduction of the column stops at 2.2$\times$10$^{25}$\,m$^{-2}$ some time
after our second observation. This final value then corresponds to the
interstellar component. If the total \ion{O}{i} column is equally
divided between CSM and ISM, the ISM may have a near-solar abundance
for oxygen, and sub-solar abundance for iron, implying that iron is
depleted by dust.

The optical flux defines the nova outburst, and thus clearly must come
from the expanding nova envelope; the ultraviolet flux changes in tandem
with the optical flux, and therefore also originates from the nova
envelope.  As can be seen from Fig.\,\ref{fig:sed}, our model fails
to predict the correct level of the ultraviolet and optical fluxes, by
several orders of magnitude.  This indicates that our model applies
only, if at all, to the X-ray part of the spectrum, which must
originate from a very different layer in the expanding envelope,
presumably much closer to the white dwarf than the layers responsible
for the optical/ultraviolet spectrum.  This conclusion is confirmed by
the observation that the marked changes in the X-ray flux during the
first XMM-Newton observation are not accompanied by changes in the
optical and/or ultraviolet fluxes. 

{Because the absorption lines from the shells are shifted from the 
rest wavelengths by the high outflow velocities, and because hot 
atmospheres do not show deep absorption lines even at the rest
wavelengths, the use of a blackbody spectrum as the central
source is not as bad as might appear at first sight.
As a first test we have tried a model atmosphere spectrum that
replaces the blackbody as the central source. Following Ness et al.~(2011) we have used the best fitting TMAP model of \citet{Rauch2010} ($007$, with log\,$g=9$ and $T=10^6$\,K)}.
This provides an even worse fit with $\chi^2$\,=\,15 {with three shells, 18 with two shells, 32 with just one shell 
and 59 without any photo-ionized absorber} (instead of 7 from the BB fit) due to the fact that \ion{O}{viii}\,/\,\ion{N}{vii} 
line ratio (the most important in the spectrum) cannot be reproduced by the atmosphere model. 
{It may be necessary to consider 
models with different abundances and outflows which, however, are 
currently not available and we therefore refrain from a systematic 
parameter study of the atmosphere model parameters.}

\section{Conclusions and prospects}
\label{sec:conclusion}

Our spectral analysis of nova V2491\,Cyg {suggests} that the absorption by highly-ionized ions
is due to {non-monotonous, possibly discrete, ejecta shells} 
with different outflow velocities and ionization levels. Variations on time scales of hours occur both in the
luminosity of the central source, and in the ionization level and
columns of the absorption shells. {We find that in our scenario of photo-ionization equilibrium 
the expanding gas is responding almost instantaneously to the variations in the source ionizing continuum. 
Our upper limits for the mass loss rate of each shell are in agreement with those estimated for other novae, especially if the shells are depleted in hydrogen.}

The values derived from the fits for the luminosity and radius of the
central source depend on the nature of the models that are applied,
and thus must be considered uncertain. In the spherically symmetric
case, however, the radius of the central emitter of X-rays decreases
monotonically with time, and thus must be in the increasingly
transparent expanding nova envelope.  The values derived for the
abundances in the ionized absorption shells do not depend on the
details of the models, and thus may be considered more secure. These
abundances indicate that the white dwarf in V2491\,Cyg is an O-Ne
white dwarf.

Improvements to our pilot models can be made in several ways.  First,
it would be useful to replace the central blackbody emitter with an appropriate white-dwarf atmosphere model.
We have tested a model in which the TMAP model substitutes the blackbody, but still absorbed by the three shells. However, this model provides results even worse than those obtained with our standard fits with a blackbody continuum.
{In the future atmosphere models must take into account different abundances and outflows.}

Second, it will be necessary to iterate
between the computation of the ionization structure of the ionizing
shells for a given spectral energy distribution, and the computation
of the spectral energy distribution and elemental abundances from the
fit, for each of the four observed spectra separately. For this
improvement it is necessary also to determine the distance more
reliably: a close distance implies a lower luminosity and thereby a
lower ionization parameter of the shells.  Finally, it would be useful
to obtain constraints on the hydrogen contents of the hot ionized
absorption shells, perhaps from ultraviolet observations. Ultraviolet
observations may also help in determining the location of origin of
the optical/ultraviolet flux.

\begin{acknowledgements}
This work is based on observations obtained with \textit{XMM-Newton}, an ESA science mission with instruments and contributions directly funded by ESA Member States and the USA (NASA). SRON is supported financially by NWO, the Netherlands Organization for Scientific Research. We also acknowledge financial support from the Faculty of the European Space Astronomy Centre. We also thank the referee for important suggestions which improved the quality of the paper.

\end{acknowledgements}

\bibliographystyle{aa}
\bibliography{bibliografia} 

\end{document}